\documentclass[runningheads]{llncs}
\usepackage{graphicx}
\usepackage{booktabs}
\usepackage[utf8]{inputenc}
\usepackage[T1]{fontenc}
\usepackage{bm}
\usepackage{amsmath,amsfonts}
\usepackage{lineno}%
\usepackage{makecell}
\usepackage{scrextend}
\usepackage{float}
\usepackage{hyperref}

\newcommand{\bi}{\begin{itemize}}
\newcommand{\ei}{\end{itemize}}
\newcommand{\ba}{\begin{array}}
\newcommand{\ea}{\end{array}}

\usepackage{xspace}

\newcommand{\bmx}[0]{\begin{bmatrix}}
\newcommand{\emx}[0]{\end{bmatrix}}

\newif\ifarxiv
\arxivtrue

\begin{document}
\title{Pitch-wide space evaluation for soccer transitions}
\titlerunning{Space evaluation at the starting point of soccer transitions}
%

\ifarxiv
\author{Yohei Ogawa\inst{1} \and Rikuhei Umemoto\inst{1} \and Keisuke Fujii\inst{1,2}}
%
\institute{
Graduate School of Informatics, Nagoya University, Nagoya, Japan. \and
Center for Advanced Intelligence Project, RIKEN, Osaka, Japan.  
\email{fujii@i.nagoya-u.ac.jp}
}
\authorrunning{Y. Ogawa et al.}
%
\else
\author{Anonymous}
\institute{}
\vspace{-11pt}

\fi
\maketitle              
\begin{abstract}
\vspace{-15pt}
Soccer is a sport played on a pitch where effective use of space is crucial. Decision-making during transitions, when possession switches between teams, has been increasingly important, but research on space evaluation in these moments has been limited. 
Recent space evaluation methods such as OBSO (Off-Ball Scoring Opportunity) use scoring probability, so it is not well-suited for assessing areas far from the goal, where transitions typically occur.
In this paper, we propose OBPV (Off-Ball Positioning Value) to evaluate space across the pitch, including the starting points of transitions. OBPV extends OBSO by introducing the field value model, which evaluates the entire pitch, and by employing the transition kernel model, which reflects positional specificity through kernel density estimation of pass distributions.
Experiments using La Liga 2023/24 season tracking and event data show that OBPV highlights effective space utilization during counter-attacks and reveals team-specific characteristics in how the teams utilize space after positive and negative transitions. 
\ifarxiv
\else
\fi

\vspace{-6pt}
\end{abstract}
\vspace{-20pt}
\section{Introduction}
\vspace{-5pt}
\label{sec:introduction}
Advances in soccer science and technology have sharpened player performance and raised the overall pace of play. Transition phases, contested at especially high speed to regain or keep possession, are now viewed as a prime expression of team style. Studies from the German Bundesliga and English Premier League indicate that transition quality strongly influences both attacking and defensive outcomes; better decision-making in these moments therefore increases the likelihood of winning \cite{Vogelbein03072014,Wright01122011}.

This study aims to identify the requirements for successful counter-attacks that arise during transitions and to characterise a transition-oriented team style. In coaching, knowing which factors enable effective counters is essential, while scouting benefits from understanding an opponent’s transition profile. Earlier work has shown that transitions shape match outcomes \cite{Reep1968,barreira2014ball} and pinpointed key components of counter-attacking play \cite{Gonzalez-Rodenas01122015}. Other studies have examined shifts from attack to defence \cite{Peters03032025,Bauer2021}. However, these investigations seldom considered the spatial configurations created by the full pitch, and transition analyses that emphasise space remain scarce (for details, see Appendix \ref{app:related}
\ifarxiv
).
\else
\footnote{Appendix is available at \url{https://anonymous.4open.science/r/OBPV_appendix}}).
\fi

Here, we present a pitch-wide space evaluation method for soccer transitions. Early research visualised pitch control with Voronoi maps \cite{taki1996development}, followed by motion-based models for player movement \cite{fujimura2005geometric,brefeld2019probabilistic}. OBSO (Off-Ball Scoring Opportunity) later provided a mathematical framework that assigns each location a likelihood of producing a goal on first touch \cite{Spearman2017,spearman2018beyond}; the model has since been transferred to other sports \cite{kono2024mathematical,iwashita2024spaceevaluationbasedpitch}. We adopt OBSO for its interpretability, but its goal-oriented design undervalues zones distant from the goal where direct scoring is unlikely.

In this paper, we propose the OBPV (Off-Ball Positioning Value) model, which enables OBSO to evaluate the entire pitch. Using this model, we can evaluate spaces such as starting points of counter-attacks based on a mathematical model that is easy to interpret. In addition, it is possible to evaluate the space during transitions, which has not been focused on in previous research \cite{Spearman2017,spearman2018beyond,kono2024mathematical}.

The contributions of this study are as follows. 
(1) The proposal of a mathematical model that evaluates space with the starting point of transitions and counter-attacks. (2) Among the modules in previous OBSO model \cite{spearman2018beyond}, the field value model was used instead of the score model to enable pitch-wide evaluation in the attack process, and the transition kernel model was used instead of the transition model to reflect the positional specificity of the pitch by estimating the kernel density from the pass distribution in each area. (3) The experiment revealed the importance of effective use of space in counter-attacks and the transition characteristics of each team in the 2023/24 La Liga season.

\vspace{-8pt}
\section{Preliminary for Space Evaluation}
\vspace{-4pt}
\label{sec:method_OBSO}
We first outline the evaluation baseline called OBSO \cite{spearman2018beyond}.  
OBSO evaluates an off-ball player by considering the joint probability  
\vspace{-3pt}
\begin{align}
    P_{OBSO}(G|D)&=\Sigma_{r\in R\times R}P(S_{r}\cap C_{r}\cap T_{r} |D) \\
    &=\Sigma_{r}P(S_{r}|C_{r}, T_{r}, D)P(C_{r}|T_{r},D)P(T_{r}|D),
\end{align}
where \(G\) denotes a goal and \(D\) represents a single frame of game data such as player positions and velocities.
\ifarxiv
 The details in OBSO are given in Appendix \ref{app:OBSO}.
\fi
$P(S_{r})$ is the probability of scoring from an arbitrary point $r\in R\times R$ on the pitch, assuming the next on-ball event occurs there.
$P(C_{r})$ is the probability that the passing team will control a ball at point $r$.
$P(T_{r})$ is the probability that the next on-ball event occurs at point $r$.
For simplicity, $P(S_{r}|D), P(T_{r}|D), P(C_{r}|D)$ are assumed to be independent if the parameter $\alpha=0$ in the original work implementation \cite{spearman2018beyond}.
Then, the joint probability can be decomposed into a series of conditional probabilities as follows:
\vspace{-3pt}
\begin{equation}
    P_{OBSO}(G|D)=\Sigma_{r\in R\times R}P(S_{r}|D)P(C_{r}|D)P(T_{r}|D).
\end{equation}
$P(C_{r}|D)$ is the probability that the attacking team will control the ball at point $r$ assuming the next on-ball event occurs there, which is called the potential pitch control field (PPCF). 
$P(T_{r}|D)$ is defined as a fixed two-dimensional Gaussian distribution with the current ball coordinates as the mean.
$P(S_{r}|D)$ is simply calculated as a value that decreases with the distance from the goal. 
We used the grid data and computed $P(C_{r}|D)$ and $P(T_{r}|D)$ according to Appendix \ref{app:OBSO}.

\begin{figure}
        \vspace{-13pt}
        \centering
        \includegraphics[width=1\linewidth]{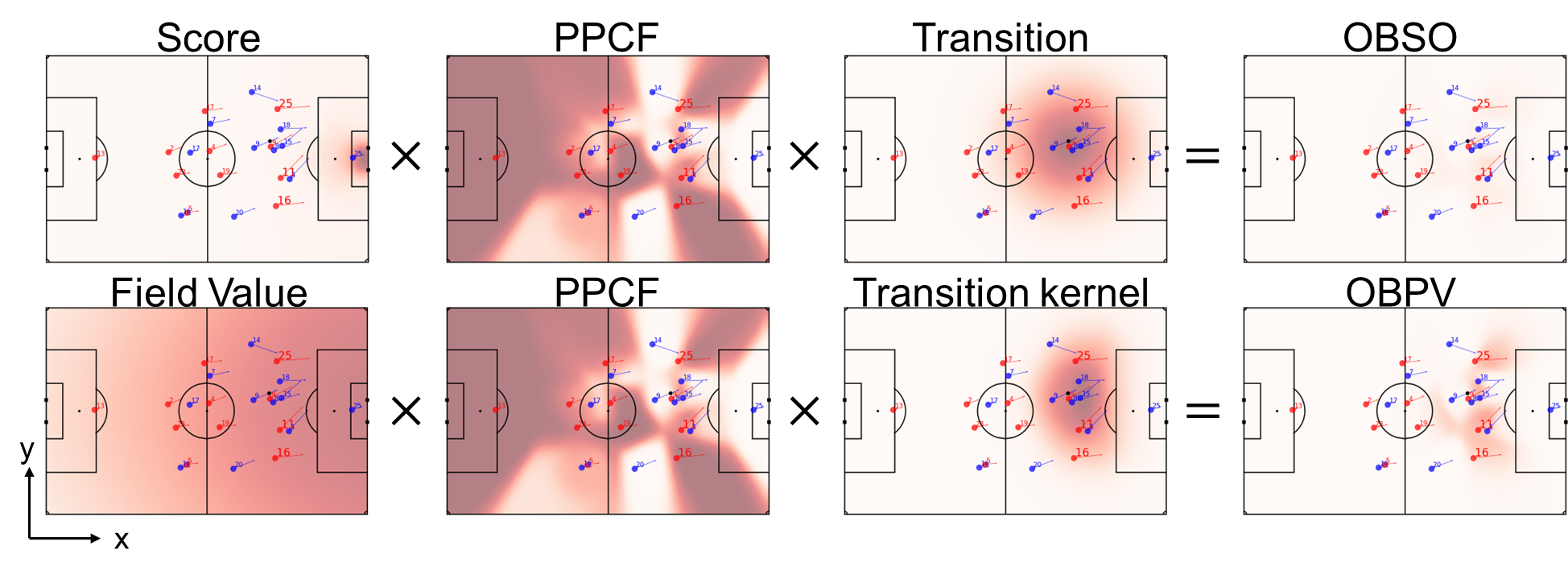}
        \vspace{-10pt}
        \caption{{\bf{Overview OBSO and OBPV.}}
            The attacking and defensive players are represented in red and blue, respectively, and the ball is represented in black. The original point (0,0) is the center of the field. One of the attacking players holds the ball near the center and attacks from left to right (x-axis). These models consist of three components: the Score model, PPCF, the Transition model for OBSO, and the field value model, PPCF, the Transition kernel model for OBPV. OBSO shows low evaluations in this pitch. OBPV allows for a more pitch-wide comprehensive assessment.
            }
        \label{fig:overview}
        \vspace{-15pt}
    \end{figure}

The overview of the OBSO \cite{spearman2018beyond} is illustrated in Fig. \ref{fig:overview} top. 
In OBSO, the scoring probability was calculated as the output $P(S_{r}|D)$ of the score model as a function of the distance from the goal. 
However, a limitation of this model is that it is designed to predict scoring opportunities, which results in generally low evaluations for positions far from the goal. To address this issue, we propose an improved model in the next section.
    
\vspace{-1pt}
\section{Proposed method}
\vspace{-3pt}
\label{sec:method_OBPV}
    In this section, we describe the proposed spatial evaluation model and the dataset used to construct it. As described in Section \ref{sec:method_OBSO}, since OBSO is a score-prediction-based metric, locations farther from the goal generally receive lower evaluations. To address this issue, we propose a pitch-wide space evaluation model called OBPV (Fig. \ref{fig:overview} bottom). 
    OBPV is composed of the field value model, which represents the importance of each location during the attacking phase; the PPCF, which is the same as that used in OBSO; and the transition kernel model, which estimates the next on-ball event location based on actual pass distributions using kernel density estimation. Through these modifications, OBPV enables spatial evaluation during attacking phases that is not limited to scoring opportunities.

    In the OBPV, the model represents the likelihood that the attacking team can complete a pass to each point on the pitch, as well as how important that location is within the context of the attack. Unlike OBSO, OBPV uses the importance of each location on the pitch as a weighted factor expressed by the following equation for a point $r$:
    \vspace{-3pt}
    \begin{equation}
        {OBPV}_{r} = w_{field} \times P(C_r|D)P(TK_r|D).
    \end{equation}
    where $w_{field}$ is represented by the field value model and $P(TK_r|D)$ is represented by the transition kernel model.
    By modeling more realistic pass destinations, the framework effectively suppresses the evaluation of players positioned behind the ball, leading to more contextually appropriate assessments.

    \vspace{-8pt}
    \subsection{Dataset}
    \vspace{-3pt}
    \label{ssec:dataset}
        Here we describe the dataset used for constructing the transition kernel model in Section \ref{ssec:TransitionKernel model}.
        We used event data provided by Statsbomb and tracking data provided by SkillCorner in Spain's top-tier division, La Liga, from the 2023/24 season.
        The event data records information on individual events such as passes and shots. It includes details such as the location of the player performing the event, the body part used to touch the ball, and information about the receiving player for passes.
        The tracking data captures the positions of all players and the ball on a frame-by-frame basis, with a frame rate of 10 fps.
        Both datasets include inherent uncertainty in the player and ball positions. Additionally, there are temporal discrepancies between the two datasets, which were resolved using a rule-based synchronization algorithm \cite{van2023etsy} that utilizes player and ball positions. Events for which synchronization failed were excluded from the evaluation in Section \ref{sec:experiments}, under the assumption that either the player or ball positions were unreliable in one of the datasets.

    \vspace{-8pt}
    \subsection{Field value model}
    \vspace{-3pt}
    \label{ssec:FieldValue model}
    In the previous score model \cite{spearman2018beyond}, locations closer to the goal are given relatively high evaluations, while areas farther from the goal are generally rated lower. As a result, it becomes unsuitable for evaluating space in situations where scoring is not the primary focus. To address this issue, we propose the field value model in our OBPV (Off-Ball Positioning Value) framework as a replacement for the Score model. The field value model considers the weight across the entire pitch, enabling spatial evaluation without solely focusing on scoring opportunities.
     
    We represent spatial importance at position $(x,y)$ by
        \begin{equation}
        \label{eq:field value}
            w_{field}(x, y) = \exp \left({-\frac{y^2}{2{\sigma(x)}^2}}\right) \times weight(x).
        \end{equation}
    The details are also given in Appendix \ref{app:fieldvaluemodel} and the coordinates and visualization are given in Fig. \ref{fig:overview} left bottom. In short, the longitudinal component $\textit{weight}(x)=\bigl(1+\exp(-(x+15)/30)\bigr)^{-1}$ is a sigmoid with the midpoint ($\textit{weight}=0.5$) at $x=-15$ m. This shape mirrors the gradual increase in attacking importance while preserving high values in the final third. 

    Lateral decay is modeled by a Gaussian whose spread narrows as play moves away from goal: $\sigma(x)=34\times{(1+\textit{weight}(x))}$. Near the penalty area the large $\sigma(x)$ keeps flanks almost as valuable as central lanes, whereas in deeper zones the reduced $\sigma(x)$ concentrates value around the center. Consequently, the field value model highlights the corridors beside the penalty box and the ``vital'' central zone, providing an interpretable, pitch-wide baseline for transition analysis.

    The differences between OBSO and OBPV, which arise from the use of the field value model, are illustrated in Fig. \ref{fig:overview} (more precisely, see Appendix Fig. \ref{fig:score_fieldvalue} using the same transition models). 
    In OBSO, which uses the Score model, the overall evaluation tends to be low. In contrast, OBPV, which incorporates the field value model, evaluates a wider variety of spaces. While it may be difficult to attempt a direct shot from the evaluated spaces, areas around player \#25 and \#16 in red suggest potential opportunities for crosses, and the space in front of player \#11 offers multiple effective options for the next play. These observations indicate that OBPV is capable of appropriately evaluating space during the attacking phase.
        
    \vspace{-8pt}
    \subsection{Transition kernel model}
    \vspace{-3pt}
    \label{ssec:TransitionKernel model}
        Here we describe the transition kernel model, one of the components of OBPV. The tendency of pass destinations varies depending on the location on the pitch. For example, passes from wide areas are more likely to be directed inward. Therefore, in this study, the transition kernel model was constructed using Kernel Density Estimation (KDE) based on actual pass distributions.
 
        KDE is one of the non-parametric methods for estimating probability density functions. In KDE, it is assumed that input variable $x_1, x_2, ..., x_n$ are independent and identically distributed, and the approximation of the probability density function is given by:
        \begin{equation}
            \hat{f}_h (x) = \frac{1}{nh}\sum^{n}_{i=1} K\left(\frac{x-x_i}{h}\right),
            \label{eq:kernel}
        \end{equation}
        where $h$ represents the bandwidth, and $K$ represents the kernel function. Although various functions can be used as kernel functions, we used Gaussian distribution according to the previous work \cite{spearman2018beyond}. 
        A prior study \cite{brefeld2019probabilistic} estimated a soccer player model by KDE, considering the player's speed and direction.
        This study demonstrated that the coverage area differs for each player position, which means that the assumption of fixed transition model of OBSO method \cite{spearman2018beyond} has a problem in particular on a full pitch.
        
        In our transition kernel model, we estimate the distribution of passes. Due to the variability in the dataset, considering ball speed and direction was deemed inappropriate and we did not use them. 
        We used the start and end positions of passes as the input variable. 
        Previous study \cite{brefeld2019probabilistic} fixed the kernel bandwidth $h$ to $0.7$, thus smoothing could not adapt to sample dispersion or sample size.  
        Here we employ Silverman’s data–driven method \cite{silverman1986density}, which selects the bandwidth by minimizing the asymptotic mean integrated squared error (AMISE). Assuming a Gaussian kernel and approximating the true density by a Gaussian, analytic minimization of the AMISE gives:
        \begin{equation}
            h_{Silverman} = \left(\frac{3}{4}n\right)^{-\frac{1}{5}} \times \sqrt{\hat{\sigma}} \times 2.
            \label{eq:silverman}
        \end{equation}
        where $\hat{\sigma}$ denotes the unbiased variance of the data, and $n$ is the number of data points.
        Thus $h_{Silverman}$ becomes narrower as $n$ increases and wider as the data variance grows, reducing the risk of under- or over-smoothing relative to a fixed bandwidth.
        The spatial distribution of pass start–end points contains sparse regions interspersed with dense clusters; our aim is to reveal overall trends rather than fine local peaks.
        In addition, we adopt a smoother estimate by doubling Silverman’s bandwidth. 
        Finally, we obtain the transition kernel model $P(TK_r) = \hat{f}_{h_{Silverman}} (x)$.

\begin{figure}[h]
            \vspace{-13pt}
            \centering
            \includegraphics[width=0.9\linewidth]{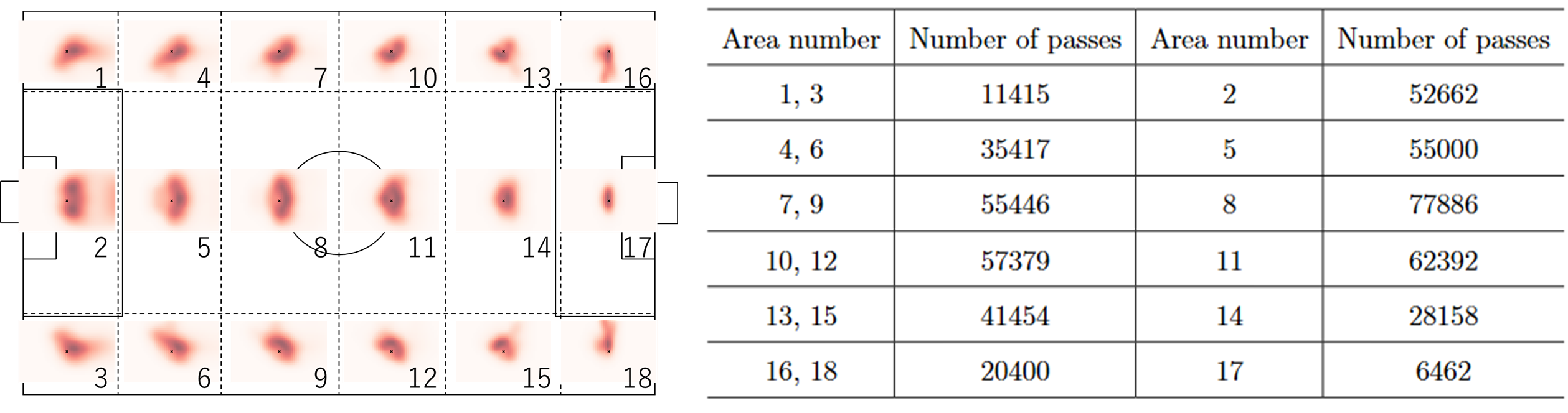}
            \vspace{-10pt}
            \caption{{\bf{The distribution of transition model and Number of passes used for kernel density estimation.
            }} 
            (Left) The distribution of transition kernel model is shown in each area. The direction of attack is to the right.
            (Right) Number of passes used for kernel density estimation is displayed. The dataset includes mirrored data, so the number of passes used is twice the actual number of passes made. In Area 17, the number is lower because players tend to opt for shooting rather than passing.
                }
            \label{fig:Transition_kernel}
            \vspace{-13pt}
        \end{figure}

        In this study, due to the data limitation, instead of modeling distributions for each individual player, we considered the distribution for each area of the pitch. The pitch division used in this study is shown in Fig. \ref{fig:Transition_kernel} left.
        The number of passes used is shown in Fig. \ref{fig:Transition_kernel} right.
        The estimated transition distributions for each area are shown in Fig. \ref{fig:Transition_kernel} left. This model successfully captures the distinctive characteristics of each area. In side areas, the distributions exhibit higher values toward the center, allowing for a more natural evaluation of player positioning. Additionally, values for backward transition are generally suppressed, resulting in relatively higher evaluations for forward spaces across the pitch.
        The example difference between the use of the transition kernel and previous transition models is illustrated on Appendix Fig. \ref{fig:gauss_kernel}. 
        
        
\vspace{-8pt}
\section{Experiments}
\vspace{-5pt}
\label{sec:experiments}
    Here we validate the effectiveness of our OBPV model via the experiments. The dataset we used is the same as the one described in Section \ref{ssec:dataset}.
    This section first examines whether OBPV discriminates between successful and failed counter-attacks, then characterizes La Liga teams’ transition profiles, and finally discusses how OBPV differs from the conventional OBSO metric.
    
    \vspace{-8pt}
    \subsection{Can OBPV explain successful counter-attacks at starting point?}
    \vspace{-3pt}
    \label{ssec:Counter_OBPV}
        First, we describe the results of the space evaluation focusing on the starting point of counter-attacks by discriminating successful and failed counter-attacks. In this study, counter-attacks that ended with a shot were defined as successful, while those that did not were defined as failed, and their use of space was compared accordingly.
        We used events labeled ``From Counter'' in the play pattern data provided by Statsbomb. Among these, only sequences in which the same team maintained possession for three consecutive events were included in the analysis. According to the definition of From Counter, three conditions must be met: (1) The possession started with an open play turnover outside the counter-attacking team’s final third. (2) The possession was at least 75\% direct towards the goal. (3) The counterattack travelled at least 18 yards towards the goal.
        We computed OBPV over the three events in each counter-attack and used the maximum OBPV within each sequence for comparison. Among the evaluated counter-attacks, 191 were classified as successful and 164 as failed.
 
        The mean values for successful and failed counter-attacks were $0.478$ and $0.426$, respectively. 
        Since normality of the two distributions was not confirmed in the results shown in Appendix Fig. \ref{fig:Counter_OBPV}, we used the Mann-Whitney U test, a non-parametric test for the comparison of the medians between two independent groups. The test yielded $p = 2.44\times10^{-6}$, confirming that successful counter-attacks had significantly higher OBPV compared to failed counter-attacks. The effect size was found to be $d = 0.243$, indicating a small to moderate effect. These findings suggest that effective space utilization may be a contributing factor to the success of counter-attacks.
        The previous OBSO \cite{spearman2018beyond} takes near zero values at the starting point of counter-attacks as shown later in Section \ref{ssec:OBSOvsOBPV}, thus the comparison between successful and failed counter-attacks would be difficult.
        
    \vspace{-8pt}
    \subsection{La Liga team transition characteristics using OBPV}
    \vspace{-3pt}
    \label{ssec:Transition_OBPV}
        Next, we show the results of the space evaluation focusing on the starting point of transition events for explaining the La Liga teams.
        The frequencies of transition events are higher than counter-attacks.
        Firstly, we describe the results of OBPV calculations at the starting point of positive transitions, which refer to transitions from defense to attack.
        Secondly, we describe those of negative transitions (from attack to defense).
        In this study, transition events were identified as sequences in which possession changed between the two teams for three consecutive events. Set pieces such as throw-ins, goal kicks, free kicks, and kick-offs after goals were excluded from the analysis.
        The average numbers of positive and negative transitions with their standard deviations for all teams are 595.85 $\pm$ 75.70 and 595.85 $\pm$ 50.25, respectively. 

        \vspace{-11pt}
        \subsubsection{OBPV at the starting point of Positive Transitions.}
            In positive transitions, we focused on transitions occurring farther from the opponent's goal, as these are considered more indicative of each team's tendencies. Therefore, transitions that occurred within 35 meters of the team's own goal were selected for evaluation. Based on the results presented in this study, the space evaluation captures how each team plays after regaining possession in deep areas. Moreover, since transitions in deeper areas are more likely to create wide open space ahead, this range is considered suitable for space evaluation.

            \begin{figure}
                \vspace{-10pt}
                \centering
                \includegraphics[width=\linewidth]{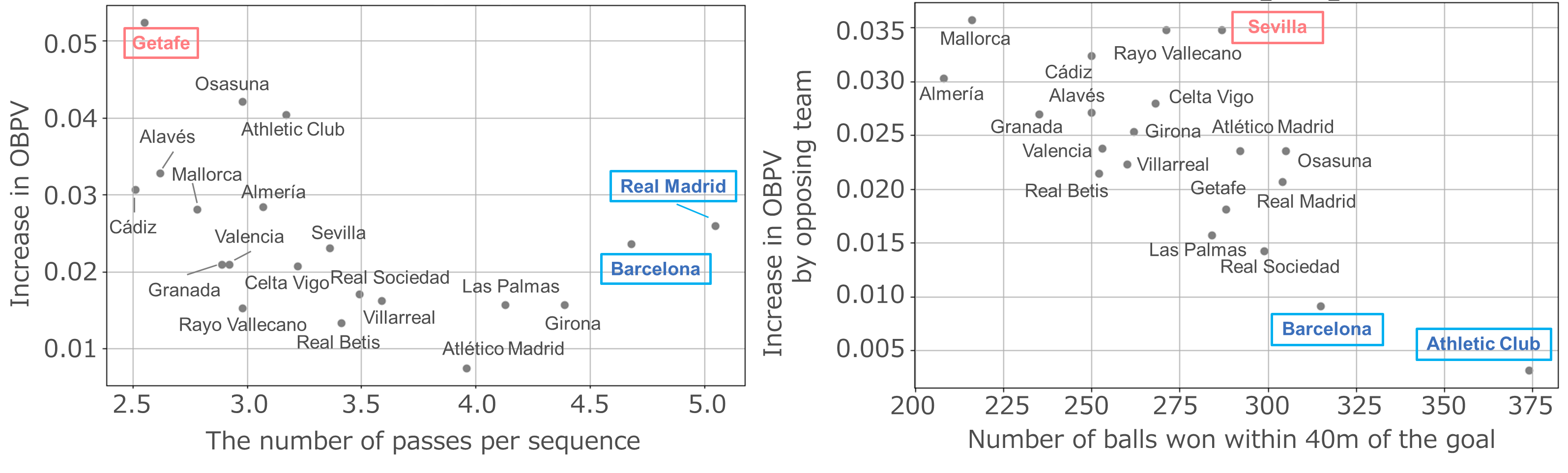}
                \vspace{-10pt}
                \caption{
                    {\bf{The correlations of OBPV increase during positive and negative transitions and other statistics}.}
                    (Left) The relationship between OBPV increase during positive transitions and the number of passes per sequence \cite{LaLigastats} is shown.
                    (Right) The relationship between the increase in OBPV during negative transitions and the number of balls won within 40 meters of the goal \cite{LaLigastats} is shown.
                }
                \label{fig:transition_corr}
                \vspace{-15pt}
            \end{figure}
            
            Figure \ref{fig:transition_corr} Left presents a scatter plot where the vertical axis represents the increase in OBPV from the first to the third event of a transition, and the horizontal axis shows the average number of passes per sequence \cite{LaLigastats}. The teams with greater OBPV increases are considered to have adopted aggressive positioning in fewer events, indicating a tendency for quick and aggressive attacks.
            Excluding Real Madrid and Barcelona, which show exceptional performances, a moderate negative correlation was observed between the two variables, with a Spearman's correlation coefficient of $r = -0.71$. This suggests that teams aiming for fast and aggressive attacks tend to lose possession more quickly.
            Real Madrid and Barcelona, despite having a higher number of passes, showed relatively large increases in OBPV, indicating that they are strong teams capable of maintaining possession even in aggressive situations. This aligns with the final standings of the 2023/24 season, where Real Madrid ranked first and Barcelona second, showing exceptional performance,  justifying their exclusion from this analysis.
            Furthermore, Getafe showed a notably higher increase in OBPV compared to other teams, confirming that they are a particularly aggressive side when launching attacks, even after regaining possession in deeper areas of the pitch.

        \vspace{-11pt}
        \subsubsection{OBPV at the starting point of Negative Transitions.}
            In negative transitions, the behavior following ball loss in advanced areas was considered to capture team tendencies better. Therefore, transitions occurring within 35 meters of the opponent's goal were selected for evaluation. Additionally, since losing possession in higher areas often creates large open spaces ahead from the opponent's perspective, this range is deemed suitable for conducting space evaluation.

            Figure \ref{fig:transition_corr} Right shows a scatter plot with the vertical axis representing the increase in OBPV from the first to the third event of a transition, and the horizontal axis representing the number of balls won within 40 meters of the goal \cite{LaLigastats}.
            Here, the OBPV reflects how much OBPV the opposing team generated against the target team. In this scatter plot, teams with a smaller increase in OBPV are considered to have executed a high-quality press that effectively prevented the opponent from exploiting space after losing the ball. A moderate negative correlation was confirmed between these variables, with a Spearman correlation coefficient of $r = -0.67$. This suggests that high-quality pressing that prevents space exploitation leads to more frequent balls won in advanced areas.
            Furthermore, both Barcelona and Athletic Club exhibited particularly low increases in OBPV, indicating that they performed excellent pressing among the 20 teams. Notably, Athletic Club also showed high OBPV increases on the left side of Figure \ref{fig:transition_corr}, suggesting they play an aggressive style of soccer both offensively and defensively.
            On the other hand, Sevilla had a relatively large increase in OBPV despite a moderate number of balls won. This may be attributed to individual performance, as Sevilla has a forward who recorded the highest number of successful tackles among forwards \cite{FBref}, suggesting that individual ability had a significant impact.

    \vspace{-8pt}
    \subsection{The difference between OBPV and OBSO}
\begin{figure}
            \vspace{-15pt}
            \centering
            \includegraphics[width=\linewidth]{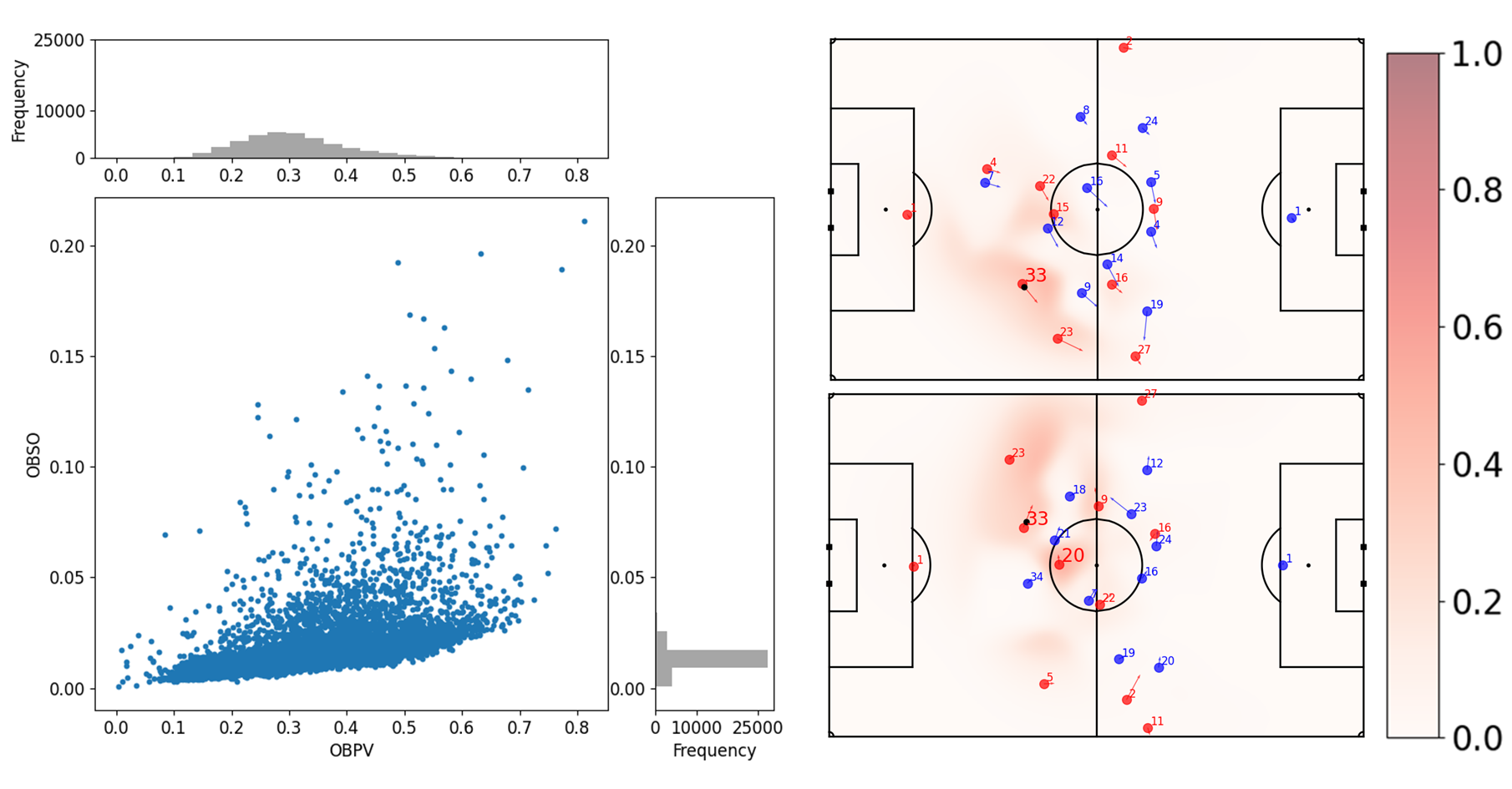}
            \vspace{-10pt}
            \caption{
                {\bf{The comparison of OBPV and OBSO, and two similar situations but different OBPVs.}}
                (Left) OBPV and OBSO value histograms for the events evaluated in Section \ref{ssec:Transition_OBPV}.  
                (Right) Two similar example situations but different OBPVs are shown. In both examples, the red team is attacking from left to right, with the blue team defending. The black circle represents the ball. 
                In these two scenes, OBSO scores were very similar: 0.01816 and 0.01817   (both red \#33). However, OBPV evaluations differed greatly: 0.29968  (upper: red \#33) and 0.50267 (lower: red \#20). 
            }
            \label{fig:OBPVvsOBSO}
            \vspace{-15pt}
        \end{figure}

    \vspace{-3pt}
    \label{ssec:OBSOvsOBPV}
        Finally, we compared the differences between OBPV and OBSO \cite{spearman2018beyond}.
        The scatter plot shown in Fig. \ref{fig:OBPVvsOBSO} Left visualizes OBPV and OBSO values calculated for the events evaluated in Section \ref{ssec:Transition_OBPV}.
        It was confirmed that the proposed OBPV enabled a broader range of evaluations for events that had received similar scores under OBSO. This was particularly evident in situations farther from the goal, which were previously given low evaluations, but where players potentially had a significant positive impact. 
         
        Examples of such positive impacts identified by OBPV are shown in Fig. \ref{fig:OBPVvsOBSO} right.
        The positioning of red \#20 in the lower example is an instance of what is called a penetrative pass target \cite{sotudeh2023potential,rahimian2022let}---a pass directed to a teammate surrounded by defenders, enabling bypassing the defensive line.
        As such, the positioning of red \#20 in the lower example is likely to serve as a crucial attacking point. This demonstrates that OBPV successfully identified the critical difference between the two examples that OBSO failed to reveal. 

\vspace{-8pt}
\section{Conclusion}
\vspace{-5pt}
\label{sec:conclusion}
    In this paper, we proposed OBPV to evaluate space across the pitch, including the starting points of transitions. 
    Experiments using La Liga season data showed that OBPV highlights effective space utilization during counter-attacks and reveals team-specific characteristics in how the teams utilize space after positive and negative transitions. 
    Future work includes incorporating factors such as interceptions during ball movement \cite{kono2024mathematical}, as well as incorporating the posture and body orientation of players, which are not available in the current dataset.

\ifarxiv
\vspace{-8pt}
\section*{Acknowledgments}
\vspace{-8pt}
This study was financially supported by JSPS KAKENHI 23H03282 and NEDO Intensive Support Program for Young Promising Researchers 24021654.
\vspace{-5pt}
\fi

\bibliographystyle{splncs04}
\bibliography{main}

\begin{thebibliography}{10}
\providecommand{\url}[1]{\texttt{#1}}
\providecommand{\urlprefix}{URL }
\providecommand{\doi}[1]{https://doi.org/#1}

\bibitem{LaLigastats}
Analyst, O.: Spanish la liga stats: 2023-24 season (2024), \url{https://theanalyst.com/2023/08/spanish-la-liga-stats-2023-24}, accessed on 14 01, 2025

\bibitem{Peters03032025}
Andrew~Peters, Nimai~Parmar, M.D., James, N.: A rule-based approach to classify counterpressing – analysis of its risks and relationship with rest defence. International Journal of Performance Analysis in Sport  \textbf{0}(0),  1--17 (2025). \doi{10.1080/24748668.2025.2473799}, \url{https://doi.org/10.1080/24748668.2025.2473799}

\bibitem{pep}
as: Pep's five-second rule, the key to city's success (2018), \url{https://en.as.com/en/2018/07/26/soccer/1532614241_079674.html}, accessed on 13 03, 2025

\bibitem{barreira2014ball}
Barreira, D., Garganta, J., Guimar{\~a}es, P., Machado, J., Anguera, M.T.: Ball recovery patterns as a performance indicator in elite soccer. Proceedings of the Institution of Mechanical Engineers, Part P: Journal of Sports Engineering and Technology  \textbf{228}(1),  61--72 (2014). \doi{10.1177/1754337113493083}

\bibitem{Bauer2021}
Bauer, P., Anzer, G.: Data-driven detection of counterpressing in professional football. Data Mining and Knowledge Discovery  \textbf{35}(5),  2009--2049 (Sep 2021). \doi{10.1007/s10618-021-00763-7}, \url{https://doi.org/10.1007/s10618-021-00763-7}

\bibitem{Bransen19}
Bransen, L., Van~Haaren, J., van~de Velden, M.: Measuring soccer players’ contributions to chance creation by valuing their passes. Journal of Quantitative Analysis in Sports  \textbf{15}(2),  97--116 (2019)

\bibitem{brefeld2019probabilistic}
Brefeld, U., Lasek, J., Mair, S.: Probabilistic movement models and zones of control. Machine Learning  \textbf{108} (01 2019). \doi{10.1007/s10994-018-5725-1}

\bibitem{Cervone2014}
Cervone, D., D’Amour, A., Bornn, L., Goldsberry, K.: Pointwise: Predicting points and valuing decisions in real time with nba optical tracking data. In: Proceedings of the 8th MIT Sloan Sports Analytics Conference, Boston, MA, USA. vol.~28, p.~3 (2014)

\bibitem{Cervone2016ASA}
Cervone, D., D’Amour, A., Bornn, L., Goldsberry, K.: A multiresolution stochastic process model for predicting basketball possession outcomes. Journal of the American Statistical Association  \textbf{111}(514),  585--599 (2016)

\bibitem{Chang2014QuantifyingSQ}
Chang, Y.H., Maheswaran, R.T., Kwok, S.J.J., Levy, T., Wexler, A.D., Squire, K.: Quantifying shot quality in the nba. In: Proceedings of the 8th MIT sloan sports analytics conference (2014), \url{https://api.semanticscholar.org/CorpusID:221170672}

\bibitem{Decroos19}
Decroos, T., Bransen, L., Van~Haaren, J., Davis, J.: Actions speak louder than goals: Valuing player actions in soccer. In: Proceedings of the 25th ACM SIGKDD Conference on Knowledge Discovery and Data Mining. pp. 1851--1861 (2019)

\bibitem{decroos2019player}
Decroos, T., Davis, J.: Player vectors: Characterizing soccer players’ playing style from match event streams. In: Joint European Conference on Machine Learning and Knowledge Discovery in Databases. pp. 569--584. Springer (2019)

\bibitem{dick2022can}
Dick, U., Link, D., Brefeld, U.: Who can receive the pass? a computational model for quantifying availability in soccer. Data Mining and Knowledge Discovery  \textbf{36}(3),  987--1014 (2022)

\bibitem{fassmeyer2021}
Fassmeyer, D., Anzer, G., Bauer, P., Brefeld, U.: Toward automatically labeling situations in soccer. Frontiers in Sports and Active Living  \textbf{3} (2021). \doi{10.3389/fspor.2021.725431}, \url{https://www.frontiersin.org/journals/sports-and-active-living/articles/10.3389/fspor.2021.725431}

\bibitem{FBref}
FBref: 2023-2024 la liga stats (2024), \url{https://fbref.com/en/comps/12/2023-2024/2023-2024-La-Liga-Stats}, accessed on 14 01, 2025

\bibitem{Fernandez18}
Fern{\'a}ndez, J., Bornn, L.: Wide open spaces: A statistical technique for measuring space creation in professional soccer. In: 12th MIT Sloan Sports Analytics Conference (2018)

\bibitem{Fernandez2019}
Fern{\'a}ndez, J., Bornn, L., Cervone, D.: Decomposing the immeasurable sport: A deep learning expected possession value framework for soccer. In: 13th MIT Sloan Sports Analytics Conference (2019)

\bibitem{fujii2025machine}
Fujii, K.: Machine learning in sports: Open approach for next play analytics (2025)

\bibitem{Fujii17}
Fujii, K., Inaba, Y., Kawahara, Y.: Koopman spectral kernels for comparing complex dynamics: Application to multiagent sport plays. In: European Conference on Machine Learning and Knowledge Discovery in Databases (ECML-PKDD'17). pp. 127--139. Springer (2017)

\bibitem{fujii2020cognition}
Fujii, K., Yoshihara, Y., Matsumoto, Y., Tose, K., Takeuchi, H., Isobe, M., Mizuta, H., Maniwa, D., Okamura, T., Murai, T., et~al.: Cognition and interpersonal coordination of patients with schizophrenia who have sports habits. PLoS One  \textbf{15}(11),  e0241863 (2020)

\bibitem{fujimura2005geometric}
Fujimura, A., Sugihara, K.: Geometric analysis and quantitative evaluation of sport teamwork. Systems and Computers in Japan  \textbf{36}(6),  49--58 (2005)

\bibitem{goes2019not}
Goes, F.R., Kempe, M., Meerhoff, L.A., Lemmink, K.A.: Not every pass can be an assist: a data-driven model to measure pass effectiveness in professional soccer matches. Big Data  \textbf{7}(1),  57--70 (2019)

\bibitem{Gonzalez-Rodenas01122015}
Gonzalez-Rodenas, J., Lopez-Bondia, I., Calabuig, F., Pérez-Turpin, J.A., Aranda, R.: The effects of playing tactics on creating scoring opportunities in random matches from us major league soccer. International Journal of Performance Analysis in Sport  \textbf{15}(3),  851--872 (2015). \doi{10.1080/24748668.2015.11868836}, \url{https://doi.org/10.1080/24748668.2015.11868836}

\bibitem{Hobbs2018QuantifyingTV}
Hobbs, J., Power, P., Sha, L., Ruiz, H., Lucey, P.: Quantifying the value of transitions in soccer via spatiotemporal trajectory clustering. In: 12th MIT Sloan Sports Analytics Conference (2018), \url{https://api.semanticscholar.org/CorpusID:221471459}

\bibitem{Liu21072015}
Hongyou~Liu, Miguel-Ángel~Gomez, C.L.P., Sampaio, J.: Match statistics related to winning in the group stage of 2014 brazil fifa world cup. Journal of Sports Sciences  \textbf{33}(12),  1205--1213 (2015). \doi{10.1080/02640414.2015.1022578}, \url{https://doi.org/10.1080/02640414.2015.1022578}, pMID: 25793661

\bibitem{iwashita2024spaceevaluationbasedpitch}
Iwashita, S., Scott, A., Umemoto, R., Ding, N., Fujii, K.: Space evaluation based on pitch control using drone video in ultimate (2024), \url{https://arxiv.org/abs/2409.14588}

\bibitem{kobayashi2023score}
Kobayashi, R., Umemoto, R., Takeda, K., Fujii, K.: Score prediction using multiple object tracking for analyzing movements in 2-vs-2 handball. In: 2023 IEEE 12th Global Conference on Consumer Electronics (GCCE). pp. 946--947. IEEE (2023)

\bibitem{kono2024mathematical}
Kono, R., Fujii, K.: Mathematical models for off-ball scoring prediction in basketball. In: ECML-PKDD workshop (2024)

\bibitem{lamas2015modeling}
Lamas, L., Santana, F., Heiner, M., Ugrinowitsch, C., Fellingham, G.: Modeling the offensive-defensive interaction and resulting outcomes in basketball. PLoS One  \textbf{10}(12),  1--14 (12 2015). \doi{10.1371/journal.pone.0144435}, \url{https://doi.org/10.1371/journal.pone.0144435}

\bibitem{link2016real}
Link, D., Lang, S., Seidenschwarz, P.: Real time quantification of dangerousity in football using spatiotemporal tracking data. PLoS One  \textbf{11}(12),  e0168768 (2016)

\bibitem{Llana20}
Llana, S., Madrero, P., Fern{\'a}ndez, J., Barcelona, F.: The right place at the right time: Advanced off-ball metrics for exploiting an opponent's spatial weaknesses in soccer. In: Proceedings of the MIT Sloan Sports Analytics Conference (2020)

\bibitem{Lucey2014}
Lucey, P., Bialkowski, A., Monfort, M., Carr, P., Matthews, I.: quality vs quantity: Improved shot prediction in soccer using strategic features from spatiotemporal data. In: Proceedings of MIT Sloan Sports Analytics Conference. pp.~1--9 (2014)

\bibitem{martens2021space}
Martens, F., Dick, U., Brefeld, U.: Space and control in soccer. Frontiers in Sports and Active Living  \textbf{3},  676179 (2021)

\bibitem{Vogelbein03072014}
Martin~Vogelbein, S.N., Hökelmann, A.: Defensive transition in soccer – are prompt possession regains a measure of success? a quantitative analysis of german fußball-bundesliga 2010/2011. Journal of Sports Sciences  \textbf{32}(11),  1076--1083 (2014). \doi{10.1080/02640414.2013.879671}, \url{https://doi.org/10.1080/02640414.2013.879671}, pMID: 24506111

\bibitem{Mchale2007}
McHale, I., Scarf, P.: Modelling soccer matches using bivariate discrete distributions with general dependence structure. Statistica Neerlandica  \textbf{61}(4),  432--445 (2007)

\bibitem{Mchale2012}
McHale, I.G., Scarf, P.A., Folker, D.E.: On the development of a soccer player performance rating system for the english premier league. Interfaces  \textbf{42}(4),  339--351 (2012)

\bibitem{mendes2024estimating}
Mendes-Neves, T., Meireles, L., Mendes-Moreira, J.: Estimating player performance in different contexts using fine-tuned large events models. arXiv preprint arXiv:2402.06815  (2024)

\bibitem{mendes2024forecasting}
Mendes-Neves, T., Meireles, L., Mendes-Moreira, J.: Forecasting events in soccer matches through language. arXiv preprint arXiv:2402.06820  (2024)

\bibitem{merckx2021measuring}
Merckx, S., Robberechts, P., Euvrard, Y., Davis, J.: Measuring the effectiveness of pressing in soccer. In: Workshop on Machine Learning and Data Mining for Sports Analytics (2021)

\bibitem{muller2021pivot}
M{\"u}ller, O., Caron, M., D{\"o}ring, M., Heuwinkel, T., Baumeister, J.: Pivot: a parsimonious end-to-end learning framework for valuing player actions in handball using tracking data. In: International Workshop on Machine Learning and Data Mining for Sports Analytics. pp. 116--128. Springer (2021)

\bibitem{narizuka2021space}
Narizuka, T., Yamazaki, Y., Takizawa, K.: Space evaluation in football games via field weighting based on tracking data. Scientific Reports  \textbf{11}(1), ~5509 (2021)

\bibitem{Pappalardo2019}
Pappalardo, L., Cintia, P., Ferragina, P., Massucco, E., Pedreschi, D., Giannotti, F.: Playerank: data-driven performance evaluation and player ranking in soccer via a machine learning approach. ACM Transactions on Intelligent Systems and Technology (TIST)  \textbf{10}(5),  1--27 (2019)

\bibitem{phatak2022}
Phatak, A., Biermann, H., Wieland, F.: Expected Counter: Probabilistic modelling of the occurrence and danger of counter-attacks in soccer before they occur. (2022)

\bibitem{Power17}
Power, P., Ruiz, H., Wei, X., Lucey, P.: Not all passes are created equal: Objectively measuring the risk and reward of passes in soccer from tracking data. In: Proceedings of the 23rd ACM SIGKDD International Conference on Knowledge Discovery and Data Mining. pp. 1605--1613 (2017)

\bibitem{rahimian2023pass}
Rahimian, P., Kim, H., Schmid, M., Toka, L.: Pass receiver and outcome prediction in soccer using temporal graph networks. In: International Workshop on Machine Learning and Data Mining for Sports Analytics. pp. 52--63. Springer (2023)

\bibitem{rahimian2022let}
Rahimian, P., da~Silva Guerra~Gomes, D.G., Berkovics, F., Toka, L.: Let’s penetrate the defense: a machine learning model for prediction and valuation of penetrative passes. In: International Workshop on Machine Learning and Data Mining for Sports Analytics. pp. 41--52. Springer (2022)

\bibitem{Raudonius2021}
Raudonius, L., Allmendinger, R.: Evaluating football player actions during counterattacks. In: Yin, H., Camacho, D., Tino, P., Allmendinger, R., Tall{\'o}n-Ballesteros, A.J., Tang, K., Cho, S.B., Novais, P., Nascimento, S. (eds.) Intelligent Data Engineering and Automated Learning -- IDEAL 2021. pp. 367--377. Springer International Publishing, Cham (2021)

\bibitem{Reep1968}
Reep, C., Benjamin, B.: Skill and chance in association football. Journal of the Royal Statistical Society. Series A (General)  \textbf{131}(4),  581--585 (1968), \url{http://www.jstor.org/stable/2343726}

\bibitem{riosneto2020new}
Rios-Neto, H., Jr., W.M., Vaz-de Melo, P.O.S.: A new look into off-ball scoring opportunity: taking into account the continuous nature of the game. In: FC Barcelona Analytics in Sports Tomorrow Congress (2020)

\bibitem{Robberechts19}
Robberechts, P.: Valuing the art of pressing. In: StatsBomb Innovation in Football Conference (2019)

\bibitem{robberechts2023xpass}
Robberechts, P., Van~Roy, M., Davis, J.: un-xpass: Measuring soccer player's creativity. In: Proceedings of the 29th ACM SIGKDD Conference on Knowledge Discovery and Data Mining. pp. 4768--4777 (2023)

\bibitem{shaw2020laurieontracking}
Shaw, L.: Laurieontracking (April 2020), \url{https://github.com/Friends-of-Tracking-Data-FoTD/LaurieOnTracking}, accessed on 11 19, 2024

\bibitem{silverman1986density}
Silverman, B.W.: Density Estimation for Statistics and Data Analysis, Monographs on Statistics and Applied Probability, vol.~26. Chapman and Hall, London (1986)

\bibitem{simpson2022seq2event}
Simpson, I., Beal, R.J., Locke, D., Norman, T.J.: Seq2event: Learning the language of soccer using transformer-based match event prediction. In: Proceedings of the 28th ACM SIGKDD Conference on Knowledge Discovery and Data Mining. pp. 3898--3908 (2022)

\bibitem{sotudeh2023potential}
Sotudeh, H.: Potential penetrative pass (p3). arXiv preprint arXiv:2302.10760  (2023)

\bibitem{spearman2018beyond}
Spearman, W.: Beyond expected goals. In: Proceedings of the 12th MIT sloan sports analytics conference. pp. 1--17 (2018)

\bibitem{Spearman2017}
Spearman, W., Basye, A., Dick, G., Hotovy, R., Pop, P.: Physics-based modeling of pass probabilities in soccer. In: Proceeding of the 11th MIT Sloan Sports Analytics Conference (2017)

\bibitem{Supola2023}
Supola, B., Hoch, T., Baca, A.: Modeling the formation of defensive gaps in basketball: Cutting on a teammate’s drive. PLOS ONE  \textbf{18}(2),  1--20 (02 2023). \doi{10.1371/journal.pone.0281467}

\bibitem{taki1996development}
Taki, T., Hasegawa, J.i., Fukumura, T.: Development of motion analysis system for quantitative evaluation of teamwork in soccer games. In: Proceedings of 3rd IEEE International Conference on Image Processing. vol.~3, pp. 815--818. IEEE (1996)

\bibitem{teranishi2022evaluation}
Teranishi, M., Tsutsui, K., Takeda, K., Fujii, K.: Evaluation of creating scoring opportunities for teammates in soccer via trajectory prediction. In: International Workshop on Machine Learning and Data Mining for Sports Analytics. pp. 53--73. Springer (2022)

\bibitem{toda2022evaluation}
Toda, K., Teranishi, M., Kushiro, K., Fujii, K.: Evaluation of soccer team defense based on prediction models of ball recovery and being attacked: A pilot study. PLoS One  \textbf{17}(1),  e0263051 (2022)

\bibitem{umemoto2023evaluation}
Umemoto, R., Fujii, K.: Evaluation of team defense positioning by computing counterfactuals using statsbomb 360 data. In: StatsBomb Conference (2023)

\bibitem{umemoto2022location}
Umemoto, R., Tsutsui, K., Fujii, K.: Location analysis of players in uefa euro 2020 and 2022 using generalized valuation of defense by estimating probabilities. arXiv preprint arXiv:2212.00021  (2022)

\bibitem{van2023etsy}
Van~Roy, M., Cascioli, L., Davis, J.: Etsy: A rule-based approach to event and tracking data synchronization. In: ECML-PKDD workshop. pp. 11--23 (2023)

\bibitem{van2021optimally}
Van~Roy, M., Robberechts, P., Davis, J.: Optimally disrupting opponent build-ups. In: Proceedings of the 2021 StatsBomb Conference, London, UK. pp. 1--16 (2021)

\bibitem{Wright01122011}
Wright, C., Atkins, S., Polman, R., Jones, B., Sargeson, L.: Factors associated with goals and goal scoring opportunities in professional soccer. International Journal of Performance Analysis in Sport  \textbf{11}(3),  438--449 (2011). \doi{10.1080/24748668.2011.11868563}, \url{https://doi.org/10.1080/24748668.2011.11868563}

\bibitem{wu2022obtracker}
Wu, Y., Deng, D., Xie, X., He, M., Xu, J., Zhang, H., Zhang, H., Wu, Y.: Obtracker: Visual analytics of off-ball movements in basketball. IEEE Transactions on Visualization and Computer Graphics  \textbf{29}(1),  929--939 (2023). \doi{10.1109/TVCG.2022.3209373}

\bibitem{yeung2024strategic}
Yeung, C., Fujii, K.: A strategic framework for optimal decisions in football 1-vs-1 shot-taking situations: An integrated approach of machine learning, theory-based modeling, and game theory. Complex \& Intelligent Systems pp. 1--20 (2024)

\bibitem{yeung2025openstarlab}
Yeung, C., Ide, K., Someya, T., Fujii, K.: Openstarlab: Open approach for spatio-temporal agent data analysis in soccer. arXiv preprint arXiv:2502.02785  (2025)

\bibitem{yeung2023transformer}
Yeung, C.C., Sit, T., Fujii, K.: Transformer-based neural marked spatio temporal point process model for football match events analysis. arXiv preprint arXiv:2302.09276  (2023)

\bibitem{Klopp}
Zeitung, S.: "wie kann es sein, dass klopp noch keinen hattrick erzielt hat?" (2015), \url{https://www.sueddeutsche.de/sport/premier-league-bei-klopps-liverpoolern-klemmt-das-gaspedal-1.2695408}, accessed on 13 03, 2025

\end{thebibliography}

\newpage
\renewcommand{\thesection}{\Alph{section}}
\appendix

\section*{Appendix}
\section{Related work}
\label{app:related}

    \subsection{Predictive analysis of team sports using tracking data}
    \label{subsec:related_anlysis}
        In many team sports, including soccer, predictive analysis using tracking data (positional information of players and ball) is widely used \cite{fujii2025machine,yeung2025openstarlab}. Cervonet et al. expressed how much scoring is expected at the end of a basketball possession as EPV (Expected Possession Value) \cite{Cervone2014,Cervone2016ASA}. This method is also used in Soccer \cite{Fernandez2019} and Handball \cite{muller2021pivot,kobayashi2023score}. 
        There have also been many predictive analyses focusing on goals, for example, Chang et al. evaluated the difficulty of shots and the ability to shoot successfully separately \cite{Chang2014QuantifyingSQ}, Lucey et al. estimated the probability of the occurrence of chances \cite{Lucey2014}, and Fujii et al. used the Koopman Spectral Kernels to predict the probability of successful shots \cite{Fujii17}.
        Furthermore, many studies evaluated players based on their scoring opportunities. McHale et al. evaluated players based on a single score without considering position and style of play \cite{Mchale2007,Mchale2012}. In contrast, Pappalardo et al. evaluated player performance in a multidimensional and role-specific way \cite{Pappalardo2019}, and Decroos and Davis used player vectors to identify players' playing styles \cite{decroos2019player}. Decroos et al. proposed VAEP (Valuing Actions by Estimating Probabilities), which evaluates player actions based on their impact on the result of the game while taking into account the previous context \cite{Decroos19}.
        However, in soccer, scoring chances are rare situations, so predictive analysis based on scoring chances is not reliable. Therefore, some researchers have focused on defense by learning ball recovery probability and probability of being attacked \cite{toda2022evaluation,umemoto2022location}, and Umemoto and Fujii evaluated player positioning in defense using counterfactuals \cite{umemoto2023evaluation}.

        Many predictive analyses that focused on each type of action have also been conducted. Power et al. evaluated passes by estimating both the risk and reward of a pass \cite{Power17}. Goes et al. evaluated passes based on whether they disrupt the opponent's defensive construction \cite{goes2019not}, and Bransen et al. evaluated passes based on how much they contributed to the xG based on similar past passing data \cite{Bransen19}.  Rahimian et al. predict and quantify Penetrative passes \cite{rahimian2023pass}, and they built a predictive model of the intended recipient of the pass and the player who actually receives it \cite{rahimian2022let}, and Robberechts et al. quantify players' creativity \cite{robberechts2023xpass}. Some researchers have also evaluated the receiver as well as the passer \cite{Llana20,fujii2020cognition,dick2022can}.
        Robberechts quantified the pressure decisions as VPEP (Valuing Pressure decisions by Estimating Probabilities), based on the VAEP \cite{Robberechts19}, and Merckx et al. automated the press analysis \cite{merckx2021measuring}. VanRoy et al. analyzed buildup tactics using Markov decision processes and described the defensive construction against buildup \cite{van2021optimally}. Furthermore, as a method for action prediction, a study based on Transformer, which is used to construct a language model, has been conducted \cite{simpson2022seq2event,yeung2023transformer,mendes2024estimating,mendes2024forecasting}. 
        In this study, the following space assessment approach is useful for evaluating player positioning for events that have no clear label and are difficult to predict, such as counter-attacks and counter-presses that occur from transitions.

    \subsection{Space evaluation for team sports}

    \label{subsec:related_offball}
        When evaluations are based on on-ball events, they often fail to assess off-ball players since only a subset of players is targeted. Moreover, off-ball time is longer than on-ball time, making it crucial to evaluate off-ball movement. Consequently, many studies have focused on evaluating off-ball movement. This evaluation has enabled the discovery of the value of players who were previously overlooked in event-based assessments.
        As a rule-based off-ball evaluation, Supola et al. evaluated off-ball cutting in basketball \cite{Supola2023}. Lamas et al. showed that off-ball movement affects the probability of getting an open shot in basketball \cite{lamas2015modeling}. Wu et al. quantitatively evaluated off-ball offensive contributions by considering player position and team cooperation \cite{wu2022obtracker}. Link et al. evaluated dangerousness of space as the probability of a player scoring a goal at each moment of ball possession \cite{link2016real}, and Fernandez evaluated moves that create space for teammates \cite{Fernandez18}. 

        Off-ball evaluations using mathematical models are also being developed. In studies considering players' dominance areas, Voronoi partitioning has been performed by treating players as generating points. Taki et al. conducted a Voronoi partitioning that calculates the minimum arrival time while considering players' speed and acceleration, evaluating space based on dominance areas \cite{taki1996development}. Additionally, regarding player motion models, Fujimura and Sugihara used a one-dimensional motion model \cite{fujimura2005geometric}, and Brefeld et al. proposed a method utilizing a model based on kernel density estimation \cite{brefeld2019probabilistic}. Furthermore, Martens et al. used a data-driven motion model \cite{martens2021space}, and Narizuka et al. used players' arrival time to weight the field \cite{narizuka2021space}.

        The model proposed in this study is an extension based on OBSO (Off-Ball Scoring Opportunity) proposed by Spearman \cite{spearman2018beyond,Spearman2017}. OBSO is a model that quantifies where a player should receive a pass to maximize scoring opportunities, using the positions of players and the ball. For more details, see Section \ref{app:OBSO}. Rios-Neto et al. extended OBSO to tracking data and provided a new perspective \cite{riosneto2020new}. Additionally, the studies by Teranishi et al. \cite{teranishi2022evaluation}, Yeung et al. \cite{yeung2024strategic}, and Umemoto and Fujii \cite{umemoto2023evaluation} each proposed models that extend OBSO. The idea of OBSO also applies to other sports, with Kono and Fujii extending it to basketball \cite{kono2024mathematical} and Iwashita et al. extending it to ultimate \cite{iwashita2024spaceevaluationbasedpitch}.

        The USO (Ultimate Scoring Opportunity) model \cite{iwashita2024spaceevaluationbasedpitch} uses pitch-wide weighting for spatial evaluation. USO calculates spatial value based on a combination of PPCF (as used in OBSO), positional weight ($w_{area}$), and distance weight ($w_{distance}$).
        In the $w_{area}$ component, spatial importance is assessed based on ease of scoring or passing, with the end zone---analogous to the goal in soccer---assigned a weight of $1$. The weights for other locations are determined by the normalized angle subtended by the end zone. 
        In contrast, the field value model proposed in this study uses distance in both the longitudinal and lateral directions rather than angles with respect to the goal. This difference stems from the structural disparity between ultimate and soccer. In soccer, using angular weighting results in extremely low values in the corners of the pitch, which is inappropriate for evaluating spatial value. Therefore, angular information was excluded from the weighting scheme, and instead, distance to the goal along each axis was used.

    \subsection{Research on transitions in soccer}

    \label{subsec:related_transition}
        In soccer, the decision-making during transitions, the moments of switching between attack and defense, has been considered important. Reep and Benjamin showed that transitions in the opponent's half have a high probability of leading to goals and significantly impact the overall performance of the team \cite{Reep1968}. On the other hand, recent studies showed the importance of counter-attacks following ball recoveries in the defensive half. Barreira et al. showed that the effectiveness of an attack increases when a team connects passes after defensive actions such as tackles in their own half \cite{barreira2014ball}. Liu et al. found that shots resulting from counter-attacks following ball recoveries in the defensive half have a significant impact on match outcomes \cite{Liu21072015}. These findings suggest that tactical changes and improvements in player skills have influenced the growing importance of ball recoveries in the defensive half, where more space is available. Wright et al. analyzed goals scored in the English Premier League and revealed that 65\% of them resulted from transitions \cite{Wright01122011}.
        Furthermore, studies have been conducted on detecting transitions and evaluating counter-attacks that originate from transitions. 
        Fassmeyer et al. utilized semi-supervised learning to detect counter-attacks and set-piece situations such as corner kicks \cite{fassmeyer2021}. Additionally, Hobbs et al. applied hierarchical clustering for transition detection and further assessed counter-attacks by calculating the defensive disorganization and the attacking threat level \cite{Hobbs2018QuantifyingTV}. Moreover, Phatak et al. introduced the concept of Expected Counter, which predicts the success rate of counter-attacks and the defensive risk level \cite{phatak2022}.
        Raudonius and Allmendinger evaluated individual players' contributions to counter-attacks using four metrics: the distance covered, the threat level of actions against opponents, the number of defenders bypassed, and the ability to control space on the field \cite{Raudonius2021}. Gonzalez-Rodenas et al. found that vertical action within the first three seconds of winning the ball increased the probability of creating scoring opportunities and was effective in creating scoring opportunities on the counter only when the opposing defense was unbalanced \cite{Gonzalez-Rodenas01122015}. 
        
        On the defensive side, Peters et al. defined Rest Defence on a rule basis, and found that when Rest Defence works, it reduces the risk of shots by opposing teams or losing possession on a counter \cite{Peters03032025}.
        Additionally, in modern soccer, the concept of counter-pressing has become an essential factor that cannot be overlooked. J\"{u}rgen Klopp, who managed Liverpool FC until 2024 and achieved great success, popularized the tactical approach known as Gegenpressing \cite{Klopp}, and Pep Guardiola, the manager of Manchester City, introduced the five-second rule in counter-pressing \cite{pep}. These managers have demonstrated success in the English Premier League, highlighting the importance of counter-pressing.
        Bauer and Anzer conducted a data-driven analysis of counter-pressing and revealed that successful counter-pressing significantly increases the goal-scoring rate, whereas failure leads to a substantial rise in conceding goals \cite{Bauer2021}. Vogelbein et al. analyzed the Bundesliga by dividing teams into three groups based on their rankings and comparing them. Their study demonstrated that the ability to quickly regain possession is a crucial factor in successful defensive performance \cite{Vogelbein03072014}.
        In this study, we focused on space, which has not been widely considered in transition evaluation, and evaluated the space at the starting point of transitions.

\section{OBSO Framework}
\label{app:OBSO}
    As described in Section 2, $P(S_{r}|D)$, $P(T_{r}|D)$ and $P(C_{r}|D)$ are represented by the Score model, PPCF (the Potential Pitch Control Field), and the Transition model, respectively. The score model is explained in Section \ref{ssec:Score model}, PPCF in Section \ref{ssec:PPCF}, and the Transition model in Section \ref{ssec:Transition model}. The overview of this model is illustrated in the upper section of Figure \ref{fig:overview}. 
    The implementation is based on the code at \cite{shaw2020laurieontracking}.
    This framework is constructed by combining the Score model, PPCF, and the Transition model, enabling goal probability estimation at different points on the pitch. However, a limitation of this model is that it is designed to predict scoring opportunities, which results in generally low evaluations for positions far from the goal. To address this issue, we propose an improved model in Section \ref{sec:method_OBPV}.

    \subsection{Score model}
    \label{ssec:Score model}
        Here we describe the Score model, which is one of the models that constitute OBSO. This model represents the importance of the pitch based on score prediction. In OBSO, it is represented as follows:
        \begin{equation}
            S(\vec{r}|\beta) = [S_d(|\vec{r}-\vec{r_g}|)]^\beta.
        \label{eq:score}
        \end{equation}
        Here, $\vec{r_g}$ denotes the position of the goal, and $S_d(x)$ refers to a monotonically decreasing function concerning distance. By setting the parameter $\beta$ , it expresses the idea that even at the same distance, scoring becomes easier if there is no defensive pressure.
        In this study, we used the grid data included in the code that implemented OBSO independently of the original authors \cite{shaw2020laurieontracking} as the Score model.

        In the Score model, locations closer to the goal are given relatively high evaluations, while areas farther from the goal are generally rated lower. As a result, it becomes unsuitable for evaluating space in situations where scoring is not the primary focus. To address this issue, we propose the field value model in our OBPV (Off-Ball Positioning Value) framework as a replacement for the Score model. The field value model considers the weight across the entire pitch, enabling spatial evaluation without solely focusing on scoring opportunities.
        
    \subsection{PPCF}
    \label{ssec:PPCF}
        Here we describe PPCF, one of the models that constitute OBSO and OBPV.
        This model represents the players' occupancy on the pitch---specifically, it can be interpreted as the probability that a player from the same team will be able to control the ball if it is passed to a given location. The control probability of player $j$ at a specific location $\vec{r}$ at time $t$ is expressed as follows:
        \begin{align}
            \frac{\textit{d}PPCF_j}{\textit{d}T}(t,\vec{r},T|s,\lambda_j) = \left( 
            1- \sum_{k} PPCF_k(t,\vec{r},T|s,\lambda_j)
            \right) f_j(t,\vec{r},T|s)\lambda_j.
            \label{eq:ppcf}
        \end{align}
        Here, $f_j(t,\vec{r},T|s)$ represents the probability that player $j$ reaches location $\vec{r}$ within time $T$. To model this, we compute the expected arrival time $\tau_{exp}(t, \vec{r})$, which represents the time required to reach a certain location $/vec{r}$, assuming the player accelerates with a constant acceleration $a$ from the initial velocity $\vec{v_j}(t)$ up to the maximum speed $v$.
        In a previous study, the values $v = 5 m/s$ and $a = 7 m/s^2$ have been used. In practice, the actual arrival time may vary due to various factors such as uncertainty in tracking data, the players' direction, awareness, and tactical decisions. To avoid modeling all these factors directly, we use the cumulative distribution function (CDF) of the logistic distribution to compute the probability that player $j$ at time $t$ can reach location $\vec{r}$ within time $T$, as shown in the following equation:
        \begin{equation}
             f_{i}(t, \vec{r}, T|s) = \frac{1}{1+\exp{\left(-\frac{T-\tau_{exp}(t, \vec{r})}{\sqrt{3}s/\pi}\right)}}.
         \label{eq:intercept}
         \end{equation}
         Here, $\sqrt{3}s/\pi$ represents the uncertainty in a player's arrival time. In this study, we set $s = 0.45$.
         $\lambda_j$ denotes the control rate. A higher value of $\lambda_j$ corresponds to a shorter time required to control the ball. This control rate is assumed to differ between the attacking and defending teams. The attacking team needs to control the ball accurately in order to shoot or make the next pass. In contrast, the defending team does not necessarily need to control the ball with high precision. To represent this distinction, we introduce the parameter $\kappa$. Using this parameter, the control rate is expressed by the following equation:
        \begin{equation}
            \lambda_{i} =
            \begin{cases} 
                \lambda & (i \in A)  \\
                \kappa\lambda & (i \in B)
            \end{cases}.
        \label{eq:lambda}
        \end{equation}
        Here, $A$ denotes the set of attacking players, and $B$ denotes the set of defending players. In a previous research, the values $\kappa=1$ and $\lambda=4.3$ were used, assuming that defending players also attempt to gain control of the ball before transitioning to attack. Based on this assumption, the same control rate was used for both attacking and defending teams in the calculations.

        Then, by integrating Equation \ref{eq:ppcf} over $T$ from $0$ to $\infty$, the PPCF for each player can be computed. Additionally, in cases where a player from one team arrives significantly earlier than any player from the opposing team and and has sufficient time to control the ball, the PPCF for that team is set to $1$, and that of the opposing team is set to $0$.

    \subsection{Transition model}
    \label{ssec:Transition model}
        Here we describe the Transition model, one of the components of OBSO. This model represents where the next on-ball event is likely to occur. In OBSO, it is assumed that the ball behaves similarly to a two-dimensional Brownian motion. This assumption is based on the idea that the ball's movement changes due to interactions such as passes, headers, blocks, and interceptions by players. Accordingly, OBSO considers a two-dimensional normal distribution. Furthermore, it is assumed that the player making the pass aims to send the ball to a location less likely to be intercepted. To reflect this, the Transition model incorporates the PPCF modeled in Section \ref{ssec:PPCF}, and is represented by the following equation:
        \begin{equation}
            T(t, \vec{r} | \sigma, \alpha) = N(\vec{r}, \vec{r_b}(t), \sigma) \cdot [\sum_{k \in A}{PPCF}_k (t, \vec{r})]^\alpha.
        \end{equation}
        In the implementation of OBSO used in this study, we set $\alpha=0$ for simplification and directly used the two-dimensional normal distribution. However, in actual football games, the tendency of pass destinations varies depending on the location on the pitch---for example, passes from the side areas are often directed inward. To account for this, the transition kernel model proposed in this study divides the pitch into 18 regions and uses kernel density estimation based on the actual pass distributions within each area.

\section{Field value model}
\label{app:fieldvaluemodel}
\begin{figure}
            \vspace{-13pt}
            \centering
            \includegraphics[width=0.7\linewidth]{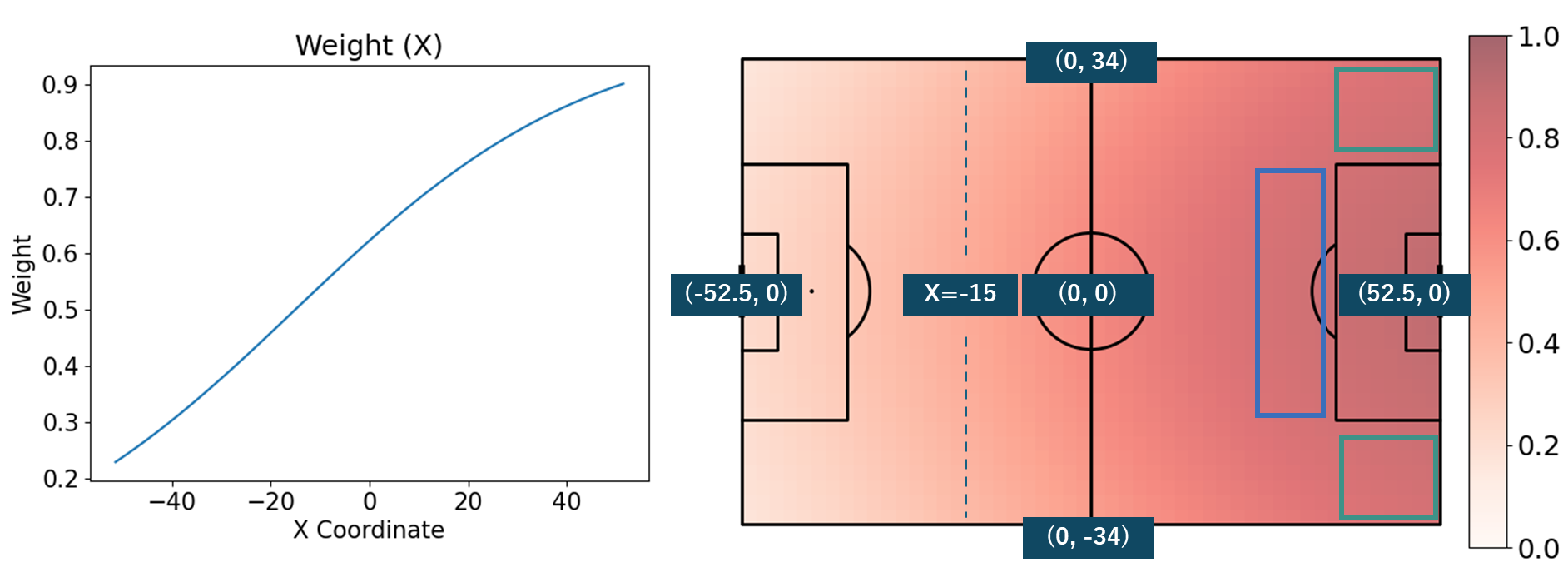}
            \caption{{\bf{Sigmoid function and field value model.}} 
            (Left) The sigmoid function is applied along the X-axis of the pitch. (Right) Field value model when attacking from left to right. Darker red indicates higher importance, while areas closer to white represent lower importance. The weight decreases as the position moves away from the goal along the length of the pitch, and also as it moves toward the sidelines. The area adjacent to the penalty box, marked in green, and the vital area, marked in blue, are assigned higher values.
                }
            \label{fig:Field Value}
        \end{figure}

        The field value model was constructed based on the following procedure. First, the following sigmoid function (Fig. \ref{fig:Field Value} Left) is defined along the longitudinal direction of the pitch (X-axis).
        \begin{equation}
        \label{eq:normmax}
            weight(x)=\frac{1}{1+\exp(-\frac{x+15}{30})}.
        \end{equation}      
        This function reflects the gradual change in importance across the pitch during an attacking phase, while maintaining high values near the goal area.
        At $X=-15$, which approximately marks the boundary between the defensive third (the third closest to the team's own goal) and the middle third of the pitch, the weight becomes $weight = 0.5$. The function is designed so that the weights are approximately $0.2$ at one end of the pitch and $0.9$ at the other. The slightly higher value than $0.2$ at the lower end is due to the additional decrease in weight as the position moves closer to the sidelines. 

        In addition, a normal distribution is applied in the width direction of the pitch (Y-axis). The maximum value of this normal distribution is given by Equation \ref{eq:normmax}, and the standard deviation is defined by Equation \ref{eq:sigma}, where the variance decreases as the position moves farther from the goal along the X-axis. This function reflects the assumption that near the goal, the importance does not significantly change even if the position shifts toward the sidelines. In the equation, the constant $34$ represents the distance from the center of the pitch to the sideline.
        \begin{equation}
        \label{eq:sigma}
            \sigma (x) = 34 \times \{1+weight (x)\} .
        \end{equation}
        Based on the above, the Field value at the position $(x, y)$ is defined as follows:
        \begin{equation}
        \label{eq:Field Value}
            w_{field}(x, y) = \exp \left({-\frac{y^2}{2{\sigma(x)}^2}}\right) \times weight(x).
        \end{equation}
        The field value model computed in this rule is illustrated on the right side of Figure \ref{fig:Field Value}.
        In the field value model, the areas adjacent to the penalty area (outlined in green) and the vital area (outlined in blue) are highly valued. These areas are considered highly important in soccer; thus, the model can be regarded as appropriately capturing their significance.

\newpage
\section{Other supplementary figures}

       \begin{figure}[h]
            \vspace{-13pt}
            \centering
            \includegraphics[width=\linewidth]
            {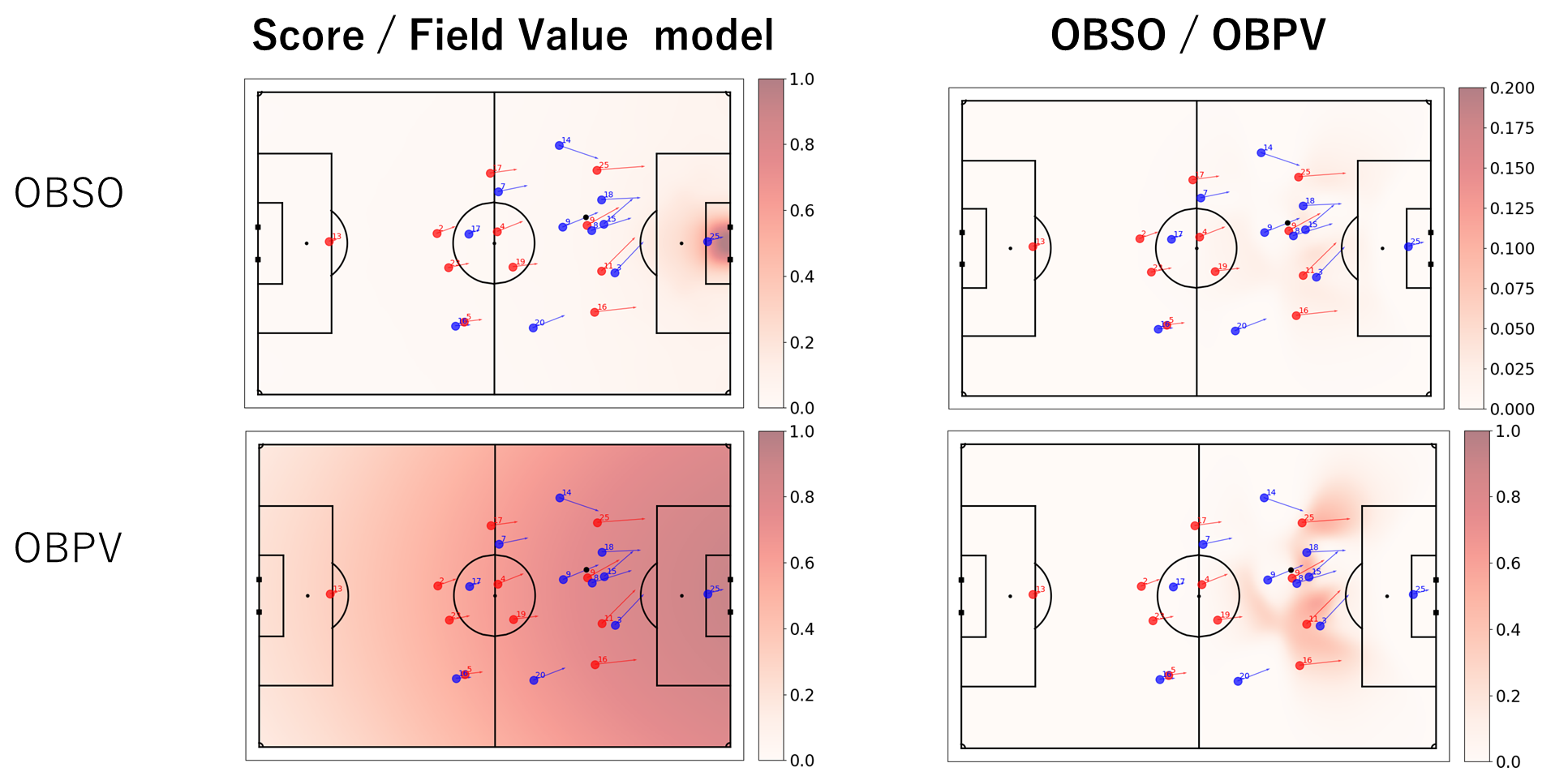}
            \caption{
                {\bf{Score model vs. Field value model}.}
                The upper row corresponds to OBSO (which uses the Score model, PPCF, and the Transition Kernel model), while the lower row corresponds to OBPV (which uses the field value model, PPCF, and the Transition Kernel model). In OBSO, the values tend to be smaller overall due to the influence of the Score model; therefore, the maximum value of the heatmap was set to $0.2$. This value is considered appropriate for scenes near the goal.
            }
            \vspace{-10pt}
            \label{fig:score_fieldvalue}
        \end{figure}

       \begin{figure}[h]
            \vspace{-13pt}
            \centering
            \includegraphics[width=\linewidth]
            {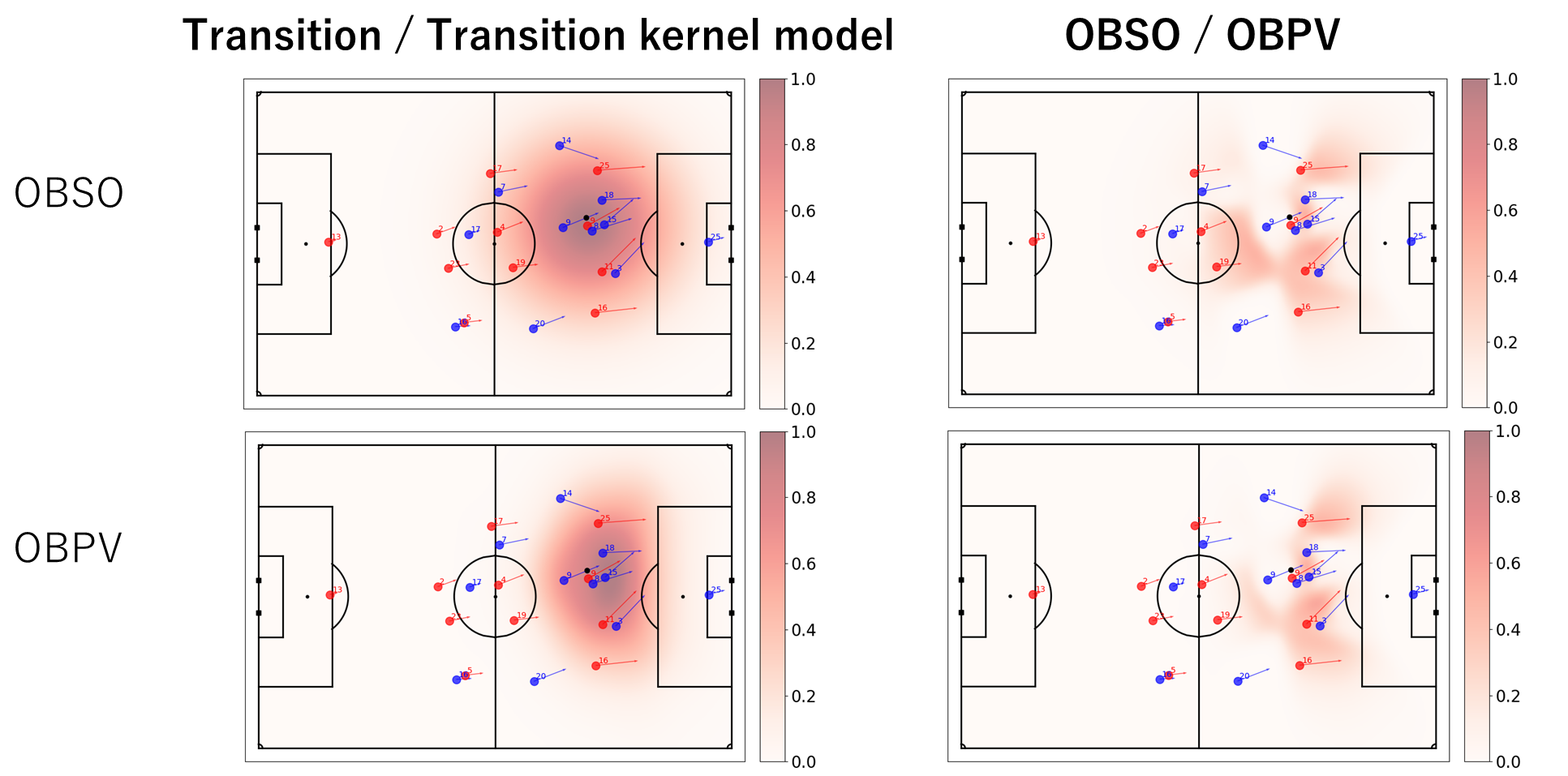}
            \caption{
                {\bf{Transition model vs. Transition kernel model}.}
                The upper row corresponds to OBSO (which uses the field value model, PPCF, and the Transition model), while the lower row corresponds to OBPV (which uses the field value model, PPCF, and the Transition kernel model).
            }
            \vspace{-10pt}
            \label{fig:gauss_kernel}
        \end{figure}

    \label{app:Counter_OBPV}

        \begin{figure}
            \vspace{-13pt}
            \centering
            \includegraphics[width=\linewidth]{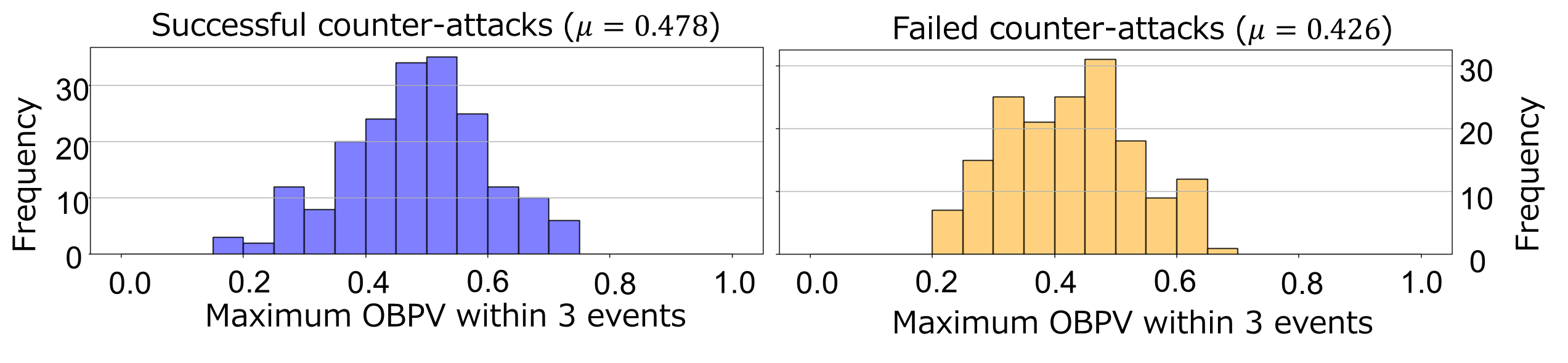}
            \caption{
                {\bf{OBPV of successful and failed counter-attacks}.}
                Successful counter-attacks showed higher OBPV values, with a statistically significant difference observed ($p < 1.0 \times 10 ^ {-5}$).
            }
            \label{fig:Counter_OBPV}
        \end{figure}


\end{document}